\documentclass[pra,twocolumn]{revtex4}
\usepackage{makeidx}
\usepackage{amssymb}
\usepackage{amsfonts}
\usepackage{amsmath}
\usepackage{graphicx}
\usepackage{bm}
\usepackage{braket}

\begin{document}

\title{Ground-state phase transitions in spin-1 Bose-Einstein condensates
with spin-orbit coupling}
\author{Xin-Feng Zhang$^{1}$}
\author{Yuan-Fen Liu$^{1}$}
\author{Huan-Bo Luo$^{1,2,3}$}
\email{huanboluo@fosu.edu.cn}
\author{Bin Liu$^{1,2}$ }
\email{binliu@fosu.edu.cn}
\author{Fu-Quan Dou$^{4}$}
\author{Yongyao Li$^{1,2}$}
\author{Boris A. Malomed$^{5,6}$}
\affiliation{$^1$School of Physics and Optoelectronic Engineering, Foshan University,
Foshan 528000, China}
\affiliation{$^2$Guangdong-Hong Kong-Macao Joint Laboratory for Intelligent Micro-Nano
Optoelectronic Technology, Foshan University, Foshan 528225, China}
\affiliation{$^3$Department of Physics, South China University of Technology, Guangzhou
510640, China}
\affiliation{$^4$College of Physics and Electronic Engineering, Northwest Normal
University, Lanzhou 730070, China}
\affiliation{$^5$Department of Physical Electronics, School of Electrical Engineering,
Faculty of Engineering, Tel Aviv University, Tel Aviv 69978, Israel}
\affiliation{$^6$Instituto de Alta Investigaci\'{o}n, Universidad de Tarapac\'{a},
Casilla 7D, Arica, Chile}

\begin{abstract}
We investigate phase transitions of the ground state (GS) of spin-1
Bose-Einstein condensates under the combined action of the spin-orbit
coupling (SOC) and gradient magnetic field. Introducing appropariate raising
and lowering operators, we exactly solve the linear system. Analyzing the
obtained energy spectrum, we conclude that simultaneous variation of the
magnetic-field gradient and SOC strength leads to the transition of excited
states into the GS. As a result, any excited state can transition to the GS,
at appropriate values of the system's parameters. The nonlinear system is
solved numerically, showing that the GS phase transition, similar to the one
in the linear system, still exists under the action of the repulsive
nonlinearity. In the case of weak attraction, a mixed state appears near the
GS transition point, while the GS transitions into an edge state under the
action of strong attractive interaction.
\end{abstract}

\maketitle


\section{Introduction}

Atomic Bose-Einstein condensates (BECs), thanks to their excellent
experimental controllability, serve as versatile platforms for simulating
and exploring various phenomena which were previously known in much more
complex form in condensed-matter physics~\cite%
{RepProgPhys.75.082401,Lewenstein}. In particular, a great deal of interest
was drawn to the BEC-enabled emulation of spin-orbit coupling (SOC), i.e.,
the effect of \ the electron spin on its motion in semiconductors. The
original SOC plays a critical role in numerous quantum phenomena and
applications, such as spin Hall effects~\cite{RevModPhys.82.1959},
topological insulators~\cite{RevModPhys.82.3045}, and the design of
spintronic devices~\cite{RevModPhys.76.323}, and more~\cite%
{liu2020holding,zhang2017spin,han2018chiral,lin2019effects,zhang2019ground,deng2012spin,Liu2022Spin, Li2013Double,Zhu2022Vortex,Zhu2022Main,lu2015spin,PRA_78_023616,FOP_Lu,FOP_YYY,FOP_YZHANG,FOP_ZRX,BEC-SOC GP eqns}%
. In the past decade, one-~\cite{nature09887}, two-~\cite%
{Science.354.83,huang2016experimental}, and three-dimensional~\cite%
{,Juzeliunas,Science.372.271} synthetic SOC in binary atomic gases have been
predicted and realized in experiments, providing pristine platforms for the
observation of novel topological phenomena. The interplay between the
synthetic SOC and intrinsic nonlinearity of ultracold atoms gives rise to a
rich variety of matter-wave states, charactized by their topological
properties and dynamical stability. These include plane waves~\cite{Han Pu},
vortices \cite{Kawakami,Drummond,
Sakaguchi,PhysRevE.89.032920,PhysRevE.94.032202,CNSNS,xu2023vortex,deng2024semi,chen2020non,guo2024stable}%
, diverse species of solitons \cite{PhysRevLett.110.264101,1D sol 2,1D sol
3,1D sol 4, Cardoso,Lobanov,2D SOC gap sol Raymond,SOC 2D gap sol
Hidetsugu,low-dim
SOC,yang2021matter,liu2024osci,wang2024bright,li2023soliton,kartashov2017solitons,saka2022one,karta2020stable}%
, and skyrmions \cite%
{PhysRevLett.109.015301,dong2022vortex,Skyrmion_ZYB,Skyrmion_LCF}. The
experimental and theoretical achievements in this field are summarized in
several reviews \cite%
{Spielman,Galitski,Ohberg,Zhai,SOC-sol-review,tu2023ryd,stepa2024spin,zhang2023anha,lyu2024super,wang2022ground,zhang2022farady,li2022quantum,xu2022three,gangwar2022dynamics,zhang2022spin,zhang2023spin,li2023bound,CSF_YZJ}%
.

The identification of the ground state (GS) in any setup, including BEC
systems, is a problem of fundamental significance. In particular, an
especially interesting possibility is to construct physically relevant
systems which exhibit GS phase transitions, transforming original excited
states into the GS as a result of variation of system's parameters. In terms
of the corresponding energy spectrum, the gap between the GS and the first
excited state closes at the phase transition point. In this vein, it has
been recently shown that in one-~\cite{PhysRevA.106.063311} and
two-dimensional~\cite{PhysRevA.109.013326} two-component SOC BEC systems (in
other words, in the SOC systems with spin 1/2) excited states with
arbitrarily high quantum numbers can be transform into the GS by adjusting
the SOC strength and the gradient magnetic field applied to the setup.
Extending that work, we here aim to reveal a possibility of GS phase
transitions in spin-1 (three-component) SOC BECs. It is natural to expect
that, in comparison to their spin-1/2 counterparts, spin-1 systems exhibit
more diverse properties \cite%
{Ma-Jia,Adhikari,He2022Multi,Wang2023Non,Lin2021Spin,FOP_Song,spin_1_PLA_QZHU,spin_1_PRA_Gangwar,spin_1_NJP_ZYP,spin_1_PRA_Rajat,spin_1_PRA_ZLC,spin_1_PRA_LLU,spin_1_PLA_JWANG}%
. In particular, in the presence of repulsive and attractive spin-spin
interactions, the GSs of the spin-1 systems are polar and ferromagnetic
states~\cite{ho}, respectively.

In this work, we introduce a spin-1 SOC\ system which includes a gradient
magnetic field and the harmonic-oscillator trapping potential. The linear
version of the system with equal SOC strength and magnetic field gradient is
solved exactly. Analysis of the respective energy spectrum reveals that
simultaneously increase of these equal parameters reduces the system's
energy, so that higher energy levels feature larger reduction rates.
Therefore, for suitable parameters, the system can undergo a GS phase
transition, with \emph{any excited state} being capable to transition into
the GS. The full nonlinear system, including repulsive or attractive
spin-spin interactions, is solved numerically. The solutions are mixed
states, in the sense that they are built, essentially, as a superposition of
several eigenstates of the linear system. In the case of relatively strong
spin-spin attraction, the nonlinear system produces narrow (self-focused)
edge states, so called because they spontaneously shift from the central
position to an edge. This finding is explained by an analytical
consideration. We analyze the results for the nonlinear system by monitoring
magnetization of the eigensates.

The following presentation is structured as follows. The model is introduced
in Section~II. The exact linear solution is reported in Section~III.
Numerical solutions of the nonlinear system with repulsive and attractive
interactions are produced in Sections.~IV and~V, respectively. The impact of
the quadratic Zeeman effect on the system is discussed in Section~VI.
Finally, the paper is concluded by Section~VII.

\section{The model}

We consider an effectively one-dimensional spin-1 (three-component) $^{87}$%
Rb BEC under the action of SOC, the trapping harmonic potential, which is
written in the scaled form as $V=x^{2}/2$, and magnetic field
\begin{equation}
\mathbf{B}=(-\alpha x,0,\Omega ),  \label{B}
\end{equation}
with constant gradient $-\alpha $ along the $x$ direction, while its $z$%
-component is the bias factor, normalized to be $\Omega =-1$. The form of
SOC is chosen as~that realized in the experiment \cite{nature09887}, i.e., $%
V_{\text{soc}}=-\beta F_{y}p_{x}$, which corresponds to an equal-weight
mixing of Rashba $(F_{x}p_{y}-F_{y}p_{x})$ and Dresselhaus $%
(-F_{x}p_{y}-F_{y}p_{x})$ couplings. Here $\beta $ is the SOC strength, and
the momentum operators are adopted in the scaled form, $\left(
p_{x},p_{y}\right) =-i\left( \partial _{x},\partial _{y}\right) $, by
setting the atomic mass and the Planck's constant to unity, i.e., $m=\hbar
=1 $. The standard set of spin-1 matrices $\mathbf{F}=(F_{x},F_{y},F_{z})$
is defined as
\begin{equation}
\begin{aligned} & F_x=\frac{1}{\sqrt{2}}\left[ \begin{array}{ccc} 0 & 1 & 0
\\ 1 & 0 & 1 \\ 0 & 1 & 0 \end{array}\right], \quad
F_y=\frac{1}{\sqrt{2}}\left[ \begin{array}{ccc} 0 & -i & 0 \\ i & 0 & -i \\
0 & i & 0 \end{array}\right], \\ & F_z=\left[ \begin{array}{ccc} 1 & 0 & 0
\\ 0 & 0 & 0 \\ 0 & 0 & -1 \end{array}\right]. \end{aligned}  \label{F}
\end{equation}%
For another form of SOC, such as $V_{\text{soc}}=\beta F_{x}p_{x}$, one may
replace magnetic field (\ref{B}) by $\mathbf{B}=(0,\alpha x,1)$, to produce the same results as
those reported here.

The single-particule Hamiltonian corresponding to what is said above is
written as
\begin{equation}
\begin{split}
\hat{H}=& \frac{p_{x}^{2}}{2}+V+V_{\text{soc}}+\mathbf{B}\cdot \mathbf{F} \\
=& -\frac{\partial _{x}^{2}}{2}+\frac{x^{2}}{2}+i\beta F_{y}\partial
_{x}-\alpha xF_{x}+\Omega F_{z}.
\end{split}
\label{H0}
\end{equation}%
In the mean-field approximation, dynamics of the spinor wave functions, $%
\Psi =(\Psi _{1},\Psi _{0},\Psi _{-1})^{T}$ of the many-body condensate is
governed by the system of Gross-Pitaevskii equations, which are presented
here in the scaled form:

\begin{equation}
\begin{split}
i\partial _{t}\Psi =& \left[ -\frac{\partial _{x}^{2}}{2}+\frac{x^{2}}{2}%
+i\beta F_{y}\partial _{x}-\alpha xF_{x}+\Omega F_{z}\right. \\
& +c_{0}\rho +c_{2}\rho \mathbf{S}\mathbf{\cdot }\mathbf{F}\bigg]\Psi ,
\end{split}
\label{main}
\end{equation}%
where $\rho =\Psi ^{\dagger }\Psi $ is the total atomic density, and the
local spin is%
\begin{equation}
\mathbf{S}=\Psi ^{\dagger }\mathbf{F}\Psi /\rho .  \label{S}
\end{equation}%
Further, coefficients $c_{0}$ and $c_{2}$ in Eq. (\ref{main}) are strengths
of the density-density and spin-spin interactions, respectively. We consider
the natural case when the former interaction is repulsive, thus fixing $%
c_{0}=1$ (unless we set $c_{0}=0$ while addressing the linear system in
Section 3), while the spin-spin interaction may have either sign.

Stationary solutions of Eq. (\ref{main}) with chemical potential $\mu $ are
sought for in the usual form,
\begin{equation}
\Psi (x,t)=\psi (x)\exp (-i\mu t),
\end{equation}%
with stationary wave functions
\begin{equation}
\psi (x)=\left( \psi _{1},\psi _{0},\psi_{-1}\right) ^{T},
\end{equation}%
satisfying the following equations:
\begin{equation}
\begin{split}
\mu \psi _{1}& =\left( -\frac{\partial _{x}^{2}}{2}+\frac{x^{2}}{2}+\Omega
+c_{0}\rho \right) \psi _{1}-(\alpha x-\beta \partial _{x})\psi _{0} \\
& +c_{2}\left[ \left( \psi _{1}\psi _{0}^{\ast }+\psi _{0}\psi _{-1}^{\ast
}\right) \psi _{0}+(|\psi _{1}|^{2}-|\psi _{-1}|^{2})\psi _{1}\right] , \\
\mu \psi _{0}& =\left( -\frac{\partial _{x}^{2}}{2}+\frac{x^{2}}{2}%
+c_{0}\rho \right) \psi _{0}-\alpha x(\psi _{1}+\psi _{-1}) \\
& +\beta \partial _{x}(\psi _{1}-\psi _{-1})+c_{2}\left( \psi _{1}^{\ast
}\psi _{0}+\psi _{0}^{\ast }\psi _{-1}\right) \psi _{1} \\
& +c_{2}\left( \psi _{1}\psi _{0}^{\ast }+\psi _{0}\psi _{-1}^{\ast }\right)
\psi _{-1}, \\
\mu \psi _{-1}& =\left( -\frac{\partial _{x}^{2}}{2}+\frac{x^{2}}{2}-\Omega
+c_{0}\rho \right) \psi _{-1}-(\alpha x+\beta \partial _{x})\psi _{0} \\
& +c_{2}\left[ \left( \psi _{1}^{\ast }\psi _{0}+\psi _{0}^{\ast }\psi
_{-1}\right) \psi _{0}-(|\psi _{1}|^{2}-|\psi _{-1}|^{2})\psi _{-1}\right] ,
\\
&
\end{split}
\label{submain}
\end{equation}%
where $\ast $ stands for the complex conjugate. The total energy of the
system is
\begin{equation}
\begin{split}
E& =\frac{1}{2}\sum_{i=\pm 1,0}\int_{-\infty }^{\infty }|\partial _{x}\psi
_{i}|^{2}dx+\frac{c_{2}}{2}\int_{-\infty }^{\infty }\rho ^{2}|\mathbf{S}%
|^{2}dx \\
& +\int_{-\infty }^{\infty }\psi ^{\dagger }\left[ \frac{x^{2}}{2}-\alpha
xF_{x}+i\beta \partial _{x}+\Omega F_{z}+\frac{c_{0}}{2}\rho \right] \psi dx.
\end{split}%
\end{equation}%
Below, we exactly solve the linear system~\eqref{submain} with $%
c_{0}=c_{2}=0 $. Then, the nonlinear system is solved numerically, for the
repulsive and attractive interactions alike.

Equations~\eqref{main} are written in the scaled form. In physical units, a
relevant value of the harmonic trapping frequency is $\omega =10$ Hz. The
number of atoms in the condensates may be $1000$, which is sufficient for
the experimental observation of the predicted patterns in full detail. The
characteristic length, time and energy are identified as $l=\sqrt{\hbar /m_{%
\mathrm{at}}\omega }=8.55$ $\mathrm{\mu }$m, $\tau =1/\omega =100$ ms, and $%
\epsilon =\hbar \omega =1.05\times 10^{-33}$ J, where $m_{\mathrm{at}%
}=1.44\times 10^{-25}$ kg is the atomic mass of $^{87}$Rb. The strength of
SOC, denoted by $\beta =\pi l/\left( \sqrt{3}\lambda \right) $, where $%
\lambda $ is the laser wavelength, can be adjusted across a wide range of
values, depending on the specific configurations of the laser system \cite%
{BEC-SOC GP eqns}. Thus, shorter wavelengths of the laser illumination makes
effective SOC strength greater. For instance, Nd:YLF lasers typically emit
light with at $\lambda =730$ nm or $\lambda =1490$ nm~\cite{laser}, which
correspond, respectively, to the SOC strength of $\beta =2.12$ and $\beta
=1.04$.

\section{The exact solution of the linear system}

Our first objective is to produce an exact solution of the linear version of
Eq.~\eqref{submain} with $\beta =\alpha $ (the balance between the SOC and
gradient magnetic field). This linear problem amounts to the form of $\mu
\psi =\hat{H}\psi $ with Hamiltonian
\begin{equation}
\hat{H}=-\frac{\partial _{x}^{2}}{2}+\frac{x^{2}}{2}-\beta (xF_{x}-i\partial
_{x}F_{y})+\Omega F_{z}.  \label{H}
\end{equation}%
To solve the linear problem exactly, we introduce an auxiliary operator,
\begin{equation}
\hat{P}=xF_{x}-i\partial _{x}F_{y}\equiv \left(
\begin{array}{ccc}
0 & \hat{a}^{\dagger } & 0 \\
\hat{a} & 0 & \hat{a}^{\dagger } \\
0 & \hat{a} & 0%
\end{array}%
\right) ,  \label{P}
\end{equation}%
where the raising and lowering operators are defined as $\hat{a}^{\dagger
}=(x-\partial _{x})/\sqrt{2}$ and $\hat{a}=(x+\partial _{x})/\sqrt{2}$. The
corresponding basis of wave functions $f_{n}(x)$ with $n\geq 0$ are provided
by eigenstates of the harmonic oscillator~\cite{LL}, \textit{viz}.,
\begin{equation}
f_{n}(x)=\left( \frac{1}{\pi }\right) ^{1/4}\frac{1}{\sqrt{2^{n}n!}}%
H_{n}(x)\exp \left( -\frac{x^{2}}{2}\right) ,  \label{f}
\end{equation}%
where $H_{n}(x)$ are the Hermite polynomials. Operators $\hat{a}^{\dagger }$
and $\hat{a}$ act on wave function~\eqref{f} according to the standard
relations,
\begin{equation}
\hat{a}^{\dagger }f_{n}(x)=\sqrt{n+1}f_{n+1}(x),\hat{a}f_{n}(x)=\sqrt{n}%
f_{n-1}(x).  \label{action}
\end{equation}

\begin{figure}[tbh]
\centering
\includegraphics[width=3.4in]{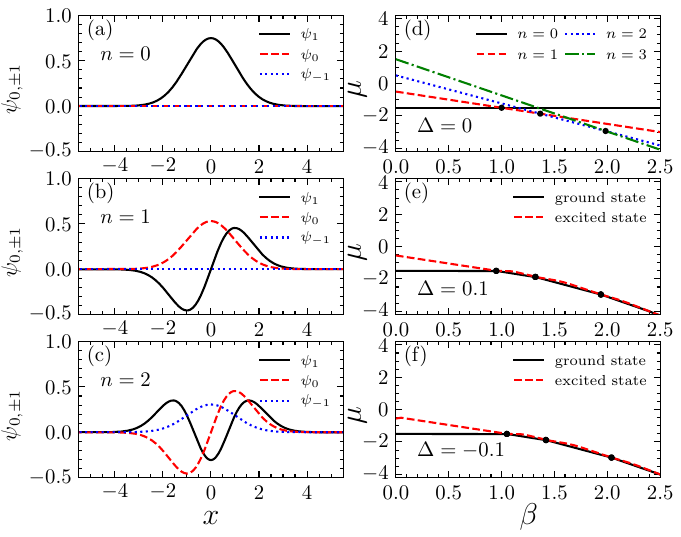}
\caption{Profiles of eigenfunctions $\protect\psi _{\pm 1,0}$, defined as
per Eqs.~\eqref{eigenstate} and \eqref{eigenstate2}, with quantum numbers
(a) $n=0$, (b) $n=1$, and (c) $n=2$. (d) The corresponding chemical
potential $\protect\mu _{n}(\protect\beta )$ (alias the corresponding values
of the energy of the linear system) in the case of $\protect\alpha =\protect%
\beta $ (i.e., $\Delta =0,$ see Eq. (\protect\ref{delta})), plotted pursuant
to Eq.~\eqref{mu_n}. The dots are values of $\protect\beta _{n}$ defined by
Eq.~\eqref{beta_n}. The chemical potentials of the GS and first excited
state, as produced by the numerical solution of Eq. (\protect\ref{coupled})
for offsets $\Delta =0.1$ and $-0.1$ are plotted in panels (e) and (f),
respectively. The dots in (e) and (f) indicate the respective GS
phase-transition points.}
\label{figure1}
\end{figure}

Note that the auxiliary operator (\ref{P}) commutes with the Hamiltonian,
i.e., $[\hat{H},\hat{P}]=0$, implying that they share the same
eigenfunctions. Solving the eigenvalue equation
\begin{equation}
\hat{P}\psi ^{(n)}=k_{n}\psi ^{(n)},  \label{EP}
\end{equation}%
we obtain the corresponding eigenvalues,
\begin{equation}
k_{n}=\left\{
\begin{array}{ll}
0, & n=0 \\
\sqrt{2n-1}, & n=1,2,3,\ldots .%
\end{array}%
\right. .  \label{k_n}
\end{equation}%
The respective eigenfunctions for $n\geq 1$ are
\begin{equation}
\psi ^{(n)}=\frac{1}{A_{n}}\left[
\begin{array}{c}
\sqrt{n}f_{n}(x) \\
\sqrt{2n-1}f_{n-1}(x) \\
\sqrt{n-1}f_{n-2}(x)%
\end{array}%
\right] ,n=1,2,3,\ldots ,  \label{eigenstate}
\end{equation}%
where the normalization coefficient $A_{n}$,%
\begin{equation}
A_{n}=\sqrt{4n-2},  \label{A_n}
\end{equation}%
secures the usual condition,
\begin{equation}
N\equiv \int_{-\infty }^{+\infty }\psi ^{\dagger }(x)\psi (x)dx=1.  \label{N}
\end{equation}
For $n=0$, the normalized wave function is
\begin{equation}
\psi ^{(0)}=\left[
\begin{array}{c}
f_{0}(x) \\
0 \\
0%
\end{array}%
\right] .  \label{eigenstate2}
\end{equation}%
The quantum number $n$ in Eqs. (\ref{EP}) - (\ref{eigenstate2}) can be used
to label the order of the eigenfunctions. Their typical profiles for $%
n=0,1,2 $ are plotted in Figs.~\ref{figure1}(a)-(c).

The eigenfunctions~\eqref{eigenstate} and (\ref{eigenstate2}) are also
eigenfunctions of the Hamiltonian $\hat{H}$, defined by Eq. (\ref{H}) with
the corresponding eigenvalues
\begin{equation}
\mu _{n}=\left\{
\begin{array}{ll}
-\frac{3}{2}, & n=0 \\
n-\beta \sqrt{2n-1}-\frac{3}{2}, & n=1,2,3,\ldots%
\end{array}%
.\right.  \label{mu_n}
\end{equation}%
It is seen that the SOC strength $\beta $, which is equal to the
magnetic-field gradient, in the case of the exact solution, alters the
spectrum~\eqref{mu_n} of eigenvalues $\mu _{n}$ but does not affect the
corresponding eigenfunctions~\eqref{eigenstate}. Figure~\ref{figure1}(d)
displays the dependence of $\mu _{n}$ on $\beta =\alpha $ for the exact
solution. Branches of $\mu _{n}$ with larger values of $n$ decrease more
rapidly as functions of $\beta $, which leads to intersections between the
branches. The intersections between the ones corresponding to $n$ and $n+1$,
which carry, alternately, the minimum energies on two sides of the
intersection points (i.e., the branches which represent the GS) directly
imply the GS phase transitions. In other words, the combined effects of
spin-orbit coupling and gradient magnetic field can close the gap between
the ground state and first excited state, thereby inducing a GS phase
transition. The corresponding critical points $\beta _{n}$ are obtained by
solving equation $\mu _{n+1}(\beta _{n})=\mu _{n}(\beta _{n}) $:
\begin{equation}
\beta _{n}=\frac{1}{2}\left( \sqrt{2n+1}+\sqrt{|2n-1|}\right)
,n=0,1,2,\ldots .  \label{beta_n}
\end{equation}%
Thus, by adjusting parameter $\beta $, the states~\eqref{eigenstate} with
\emph{any quantum number} $n$ can be transformed into the GS.

A basic principle of quantum mechanics is that, in the framework of the
single Schr\"{o}dinger equation, the GS cannot be degenerate \cite{LL}. In
the present case, the system of three coupled Schr\"{o}dinger-like equations
(\ref{submain}) violates this principle at critical points (\ref{beta_n}),
as at these points two different eigenstates, corresponding to the
intersecting branches $\mu (\beta )$, simultaneously represent the GS.

The analytical results were presented under the solvability condition $%
\alpha =\beta $. In the general case,
\begin{equation}
\alpha =\beta +\Delta ,  \label{delta}
\end{equation}%
with offset $\Delta $ between the magnetic field gradient $\alpha $ and SOC
strength $\beta $, the linear system does not admit analytical solutions,
but it can be solved approximately. To this end, solutions are looked for as
finite combination of the harmonic-oscillator wave functions~\eqref{f}
truncated at $n=N_{t}$ :
\begin{equation}
\begin{split}
\psi _{1}& =\sum_{n=0}^{N_{t}}b_{n}f_{n}(x), \\
\psi _{0}& =\sum_{n=0}^{N_{t}}e_{n}f_{n}(x), \\
\psi _{-1}& =\sum_{n=0}^{N_{t}}g_{n}f_{n}(x),
\end{split}
\label{ansatz}
\end{equation}%
where $b_{n}$, $e_{n}$ and $g_{n}$ are coefficients to be determined. Here,
we produce results for $N_{t}=50$, which provides practically exact results.
Substituting the ansatz~based on Eqs. \eqref{delta} and \eqref{ansatz} in
the linearization of Eq.~\eqref{submain}, we obtain a set of coupled linear
equations for $b_{n}$, $e_{n}$ and $g_{n}$:
\begin{equation}
\begin{split}
2\mu b_{n}=& (2n-3)b_{n}-\Delta \sqrt{n+1}e_{n+1}-(2\beta +\Delta )\sqrt{n}%
e_{n-1}, \\
2\mu e_{n}=& -\Delta \sqrt{n}b_{n-1}-(2\beta +\Delta )\sqrt{n+1}%
b_{n+1}+(2n-1)e_{n} \\
& -(2\beta +\Delta )\sqrt{n}g_{n-1}-\Delta \sqrt{n+1}g_{n+1}, \\
2\mu g_{n}=& (2n+1)g_{n}-\Delta \sqrt{n}e_{n-1}-(2\beta +\Delta )\sqrt{n+1}%
e_{n+1}.
\end{split}
\label{coupled}
\end{equation}

Equations~\eqref{coupled} can be solved by numerically by diagonalizion of
the corresponding matrix. The corresponding GS phase transitions occur at
intersection of the two lowest-energy branches, as shown in Figs.~\ref%
{figure1}(e) and (f), which correspond to offsets $\Delta =0.1$ and $-0.1$,
respectively, in Eq. (\ref{delta}). The results clearly indicate that the GS
phase transition occurs at $\alpha \neq \beta $, similar to what is found
above in the exactly solvable linear system with $\alpha =\beta $.

\section{Numerical results for the system with repulsive spin-spin
interactions}

Next, we address the spin-1 system in the full form, which is based on Eq.~%
\eqref{main} including the nonlinear terms. To facilitate the comparison of
nonlinear bound states with their linear counterparts, we here keep the same
constraint $\alpha =\beta $ which was imposed above to produce the exact
solution of the linear system. In this case, the system's GS can be found in
a numerical form by means of the imaginary-time propagation method \cite%
{chiofalo,Bao}. In this context, we fix the total norm as $N=1$, see Eq. (%
\ref{N}). We used a $512$-point grid to discretize the space, with $x$
ranging in the interval of $[-8,+8]$. The spatial and time derivatives were
handled by means of the Fourier transform and backward Euler scheme,
respectively. The results definitely converged after $\sim 10^{5}$ steps.

We start the consideration of the nonlinear system with one featuring the
repulsive spin-spin interaction, i.e., $c_{2}\geq 0$ in Eq.~\eqref{main}.
The term in the system's energy accounting for the spin-spin interaction is
\begin{equation}
E_{s}=\frac{c_{2}}{2}\int_{-\infty }^{+\infty }\rho ^{2}|\mathbf{S}|^{2}dx
\label{Es}
\end{equation}%
(recall $\rho $ is the total atomic density, and $\mathbf{S}$ is the local
spin defined as per Eq. (\ref{S})).\ To minimize the energy, local spins
tend to evolve towards $\mathbf{S}=0$, which means the system will try to
construct a polar state, with zero magnetization. Typical spinors
corresponding to the polar states are
\begin{equation}
\xi _{1}=\frac{1}{\sqrt{2}}\left[
\begin{array}{c}
1 \\
0 \\
1%
\end{array}%
\right] ,\xi _{2}=\left[
\begin{array}{c}
0 \\
1 \\
0%
\end{array}%
\right] ,  \label{xi}
\end{equation}%
where $\xi \equiv \psi /\sqrt{\rho }$, hence the local spin vector (\ref{S}%
)\ is written as $\mathbf{S}=\xi ^{\dagger }\mathbf{F}\xi $. In the polar
state $\xi _{1}$, particles are evenly distributed between the $\psi _{1}$
and $\psi _{-1}$ components. In the polar state $\xi _{2}$, all particles
populate solely the $\psi _{0}$ component, hence the polar states always
have $\psi _{-1}=\psi _{1}$.

\begin{figure}[h]
\centering
\includegraphics[width=3.4in]{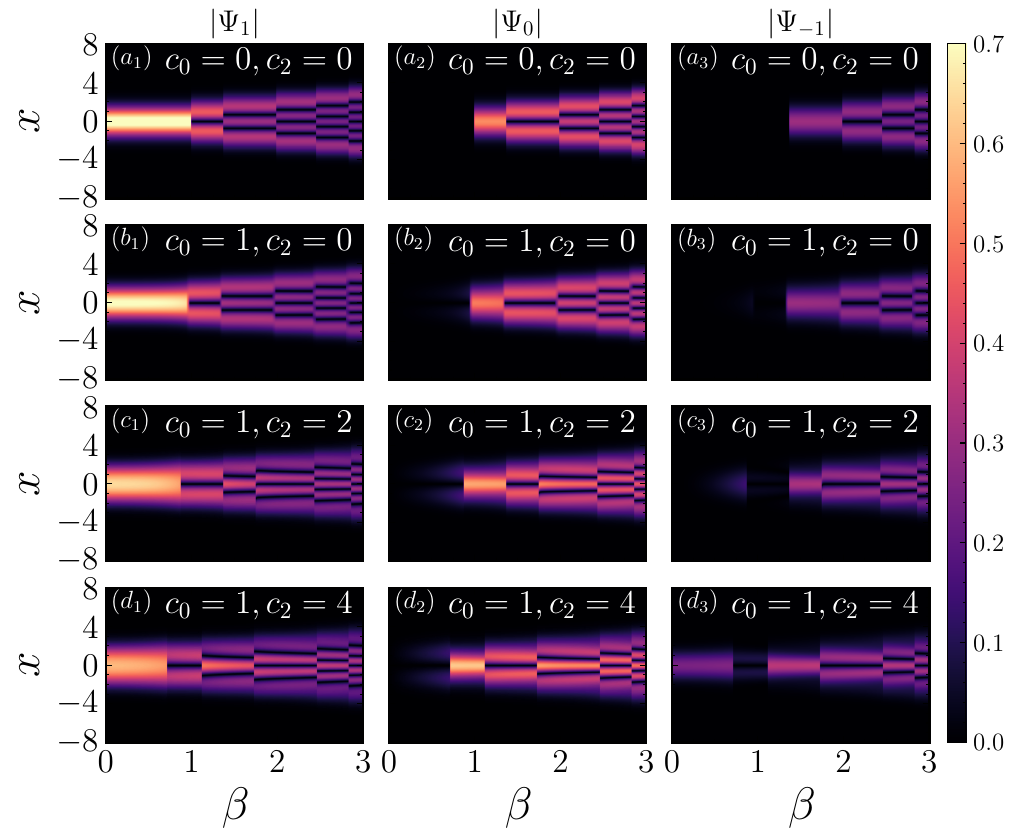}
\caption{Distributions of the absolute values of the wave functions in the
three components of the ground state, plotted in panels $a_{1-3}$ as per the
exact linear solution (\protect\ref{eigenstate}) and critical points (\protect
\ref{beta_n}), and in other panels as produced by the numerical solution.
The parameters are $c_0=c_2=0$ in panels $a_{1-3}$; $c_0=1,c_2=0$ in $%
b_{1-3} $, $c_0=1,c_2=2$ in $c_{1-3}$, and $c_0=1,c_2=4$ in $d_{1-3}$.}
\label{figure2}
\end{figure}

In the general case, the magnetization vector of the spin-1 BEC is defined
as~\cite{wenlin}
\begin{equation}
\mathbf{M}=\int_{-\infty }^{\infty }\rho \mathbf{S}dx\equiv \int_{-\infty
}^{\infty }\psi ^{\dagger }\mathbf{F}\psi dx  \label{M}
\end{equation}
(recall $\mathbf{F}$ is the matrix vector (\ref{F})), which obviously
vanishes in the polar states. For linear solutions~\eqref{eigenstate} and~%
\eqref{eigenstate2}, $\mathbf{M}$ takes values
\begin{equation}
M_{x}=0,M_{y}=0,M_{z}=\left\{
\begin{array}{ll}
1, & n=0 \\
\frac{1}{4n-2}, & n=1,2,3,\ldots%
\end{array}%
.\right.  \label{Mz}
\end{equation}

Figures \ref{figure2} ($b_{1-3}$), ($c_{1-3}$),and ($d_{1-3}$) display maps
of the distribution of absolute values of the three spinor components, $%
\left\vert \Psi _{+1,0,-1}\right\vert $ in the numerically found GS for $%
\beta =\alpha $ ranging from $0$ to $3$, and $c_0=1,c_{2}=0,2,4$. For the
comparison's sake, similar plots for the exact linear solution, with $c_0 =
c_2 = 0$, are displayed in Figs.~\ref{figure2}($a_{1-3}$). This is
consistent with what is predicted by the exact linear solution~%
\eqref{eigenstate}. Note also that, with the increase of $n $, the GS
pattern develops a striped structure, which is a generic feature of SOC
systems~\cite{nature09887,stripe1,stripe2}. According to the linear solution~%
\eqref{eigenstate}, we can find that the wave functions of the three
components are solutions of harmonic potential with adjacent quantum
numbers. Due to the symmetry of the harmonic potential, the solutions
alternate between odd and even parity as the quantum number increases. This
leads to complementary properties between adjacent solutions, as shown in
Figs.~\ref{figure2}.

\begin{figure}[th]
\centering
\includegraphics[width=3.4in]{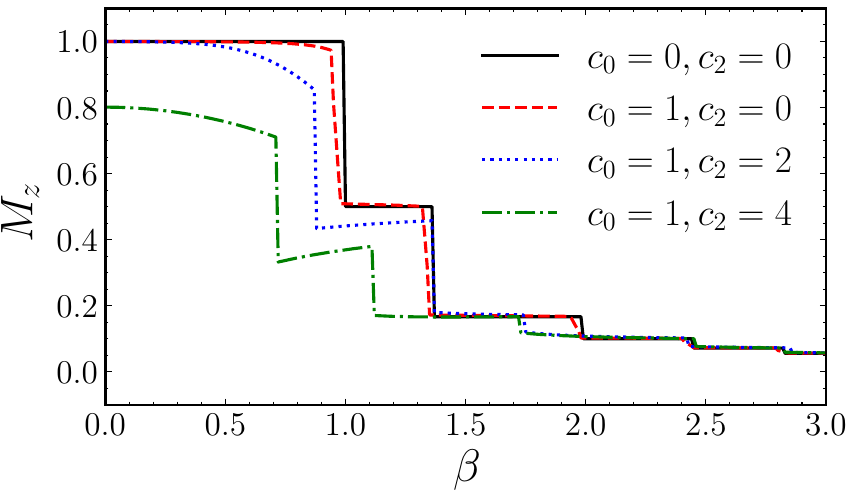}
\caption{The magnetization of the system's GS vs. the equal SOC\ strength
and magnetic-field gradient for fixed values of the coefficient of the
repulsive interaction, \textit{viz}., $c_0=1$ and $c_{2}=0,2,4$.
Additionally, $c_0=c_2=0$ corresponds to the linear case.}
\label{figure3}
\end{figure}

As the strength $c_{2}$ of the spin-spin repulsion increases, all critical
points $\beta _{n}$ of the GS phase transition shift towards $\beta =0$.
This happens because, as mentioned above, the increase of $c_{2}$ drives all
the system's eigenstates closer to polar ones, hence a smaller difference
between them needs a smaller SOC strength to perform the GS phase transition.

To produce a clear representation of the impact of the spin-spin interaction
parameter $c_{2}$ on the GS magnetization $M_{z}$, Fig.~\ref{figure3}
depicts it as a function of $\beta $ for different fixed values of $c_{2}$.
It is seen that, in accordance with the above-mentioned trend of the
suppression of the magnetization by the repulsive spin-spin interaction, $%
M_{z}$ decays with the increase of $c_{2}$. Using Eq.~\eqref{Mz}, we obtain
the decay rate of the magnetization with the increase of $n$:
\begin{equation}
\Delta M_{z}=M_{z}(n+1)-M_{z}(n)=-\frac{1}{4n^{2}-1},n=0,1,2,\ldots
\label{dM}
\end{equation}%
This result implies that, as $n$ increases, the decay of $M_{z}$ weakens,
which is consistent with what is shown in Fig.~\ref{figure3}.

The structure of the numerically found GS in can be analyzed if it
represented by a superposition of the linear eigenstates~\eqref{eigenstate},
\begin{equation}
\psi (x)=\sum_{n=0}^{N_s }d_{n}\psi ^{(n)}(x),  \label{expand}
\end{equation}%
where $|d_{n}|^{2}$ accounts for the weight of each eigenstate, satisfying
the normalization condition, $\sum_{n=0}^{N_t }|d_{n}|^{2}=1$, which is a
corollary of the unitary normalization adopted for $\psi (x)$. We used $N_s
= 20$ in our calculations, which efficiently provides sufficiently accurate
results.

\begin{figure}[h]
\centering
\includegraphics[width=3.4in]{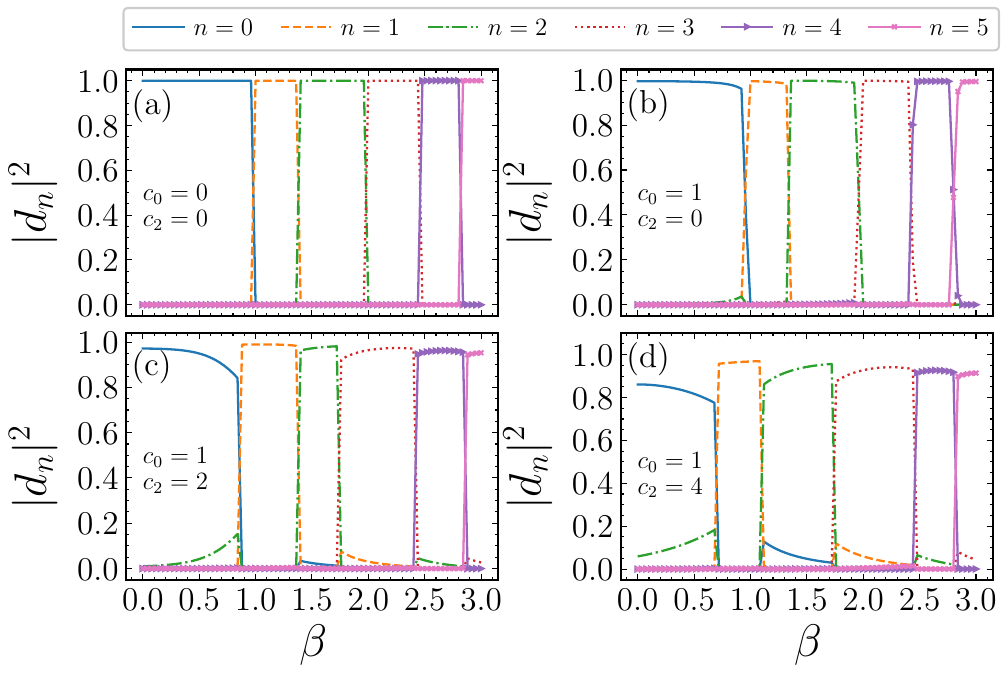}
\caption{The dependence of weights $|d_{n}|^{2}$ of linear eigenstates in
the GS expansion~\eqref{expand} on $\protect\beta $ for (a) $c_0=c_{2}=0$,
(b) $c_0=1,c_{2}=0$, (c) $c_0=1,c_{2}=2$ and (d) $c_0=1,c_{2}=4$. Panels (c)
and (d) demonstrate that the system's GS becomes a superposition of a pair
of next-nearest-neighbor eigenstates, labeled by quantum number $n$ and $n+2$%
.}
\label{figure4}
\end{figure}

Figure~\ref{figure4} shows the dependence of weights $|d_{n}|^{2}$ on $\beta
$ for (a) $c_{0}=c_{2}=0$, (b) $c_{0}=1,c_{2}=0$, (c) $c_{0}=1,c_{2}=2$ and
(d) $c_{0}=1,c_{2}=4$. Figure~\ref{figure4}(a) corresponds to the linear
solution, where the weights of the two eigenstates with adjacent quantum
numbers, i.e., $n$ and $n+1$, exhibit a discontinuous change at the GS phase
transition point. For example, at the first phase transition point where $%
\beta _{0}=1$, the weight $|d_{0}|^{2}$ decreases from 1 to 0, while the
weight $|d_{1}|^{2}$ increases from 0 to 1. Figure~\ref{figure4}(b)
corresponds to $c_{0}=1,c_{2}=0$, and we can see that the pattern is similar
to the one in the linear system. As $c_{2}$ increases, Figs.~\ref{figure4}%
(c) and (d) demonstrate that the system's GS becomes a superposition of a
pair of next-nearest-neighbor eigenstates, labeled by quantum number $n$ and
$n+2$. As previously mentioned, although the phase transition point has
shifted closer to $\beta =0$, it can still be easily identified in Fig.~\ref%
{figure4}.

\section{Numerical results for the system with attractive spin-spin
interactions}

\begin{figure}[tbph]
\centering
\includegraphics[width=3.4in]{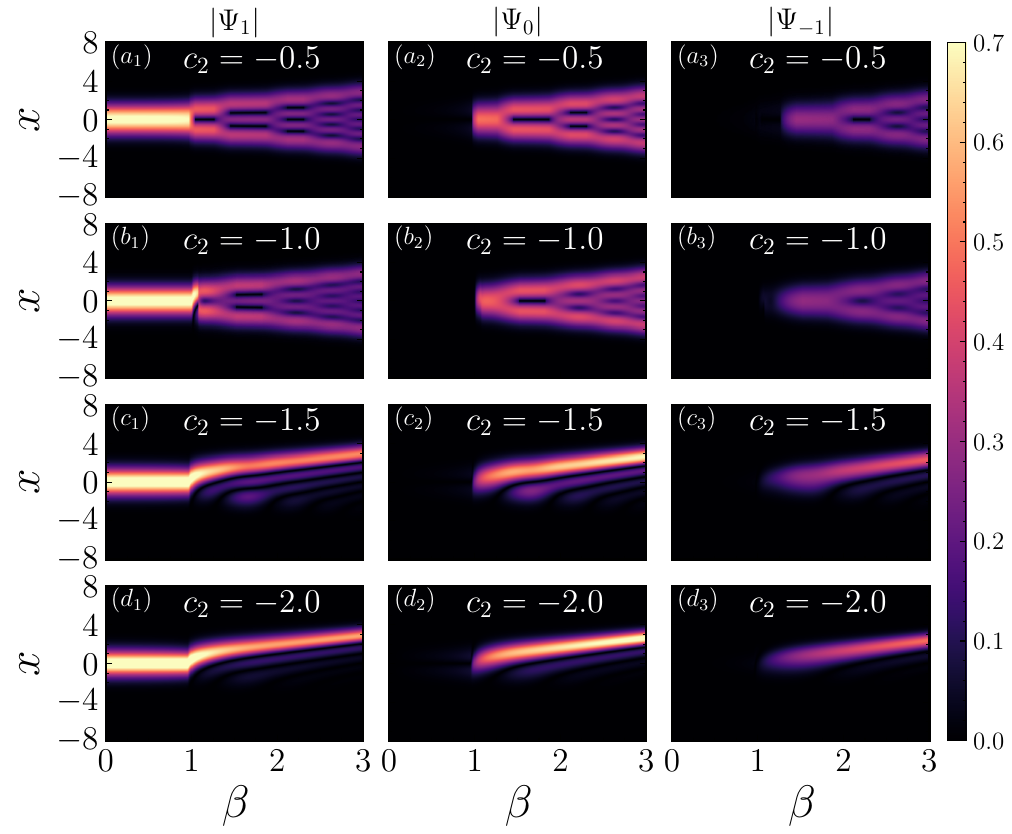}
\caption{The same as in Fig. \protect\ref{figure2}, but in the case of the
attractive spin-spin interaction, \textit{viz}., for $c_{2}=-0.5$ ($%
a_{1}-a_{3}$), $c_{2}=-1.0$ ($b_{1}-b_{3}$), $c_{2}=-1.5$ ($c_{1}-c_{3}$),
and $c_{2}=-2.0$ ($d_{1}-d_{3}$). The density-density interaction remains
repulsive, with $c_{0}=1$. As $c_2$ decreases, the system's GSs transition
into edge states, as shown in panels ($c_{1-3}$) and ($d_{1-3}$).}
\label{figure5}
\end{figure}

Next, we consider the nonlinear system with the attractive spin-spin
attraction, i.e., $c_{2}<0$ in Eq.~\eqref{main}. Figure~\ref{figure5}
displays maps of the distribution of absolute values of the three spinor
components, $\left\vert \Psi _{+1,0,-1}\right\vert $ in the numerically
found GS for $\beta =\alpha $ ranging from $0$ to $3$, and $%
c_{2}=-0.5,-1.0,-1.5,-2.0$, cf. Fig. \ref{figure2}. The figure exhibits
completely different patterns for different values of $c_{2}$. At $%
c_{2}=-0.5 $ and $-1$, we observe the presence of spatially symmetric GS,
with $\left\vert \psi_{j}(-x)\right\vert =\left\vert \psi_{j}(x)\right\vert $%
, in Figs.~\ref{figure5}($a_{1-3}$) and ($b_{1-3}$). For the stronger
spin-spin attraction, \textit{viz}., $c_{2}=-1.5$ and $-2$, the GS take the
form of spatially narrow edge states, strongly shifted to $x>0$, at $\beta
>1 $. The narrowness of the GS in this case is explained by the
self-focusing effect of the attractive nonlinearity. The shift from the
central position towards $x<0$, which is explained analytically below, is
possible, as Eq. (\ref{main}) is not symmetric wih respect to the
reflection, $x\rightarrow -x $. The symmetry breaking is characterized by
the dependence of $\bar{x}$ on $\beta$, as shown in Fig.~\ref{figure7}(b),
where $\bar{x}=\int_{-\infty}^{\infty}\psi^\dagger x\psi dx$ is the average
displacement.

\begin{figure}[tbp]
\centering
\includegraphics[width=3.4in]{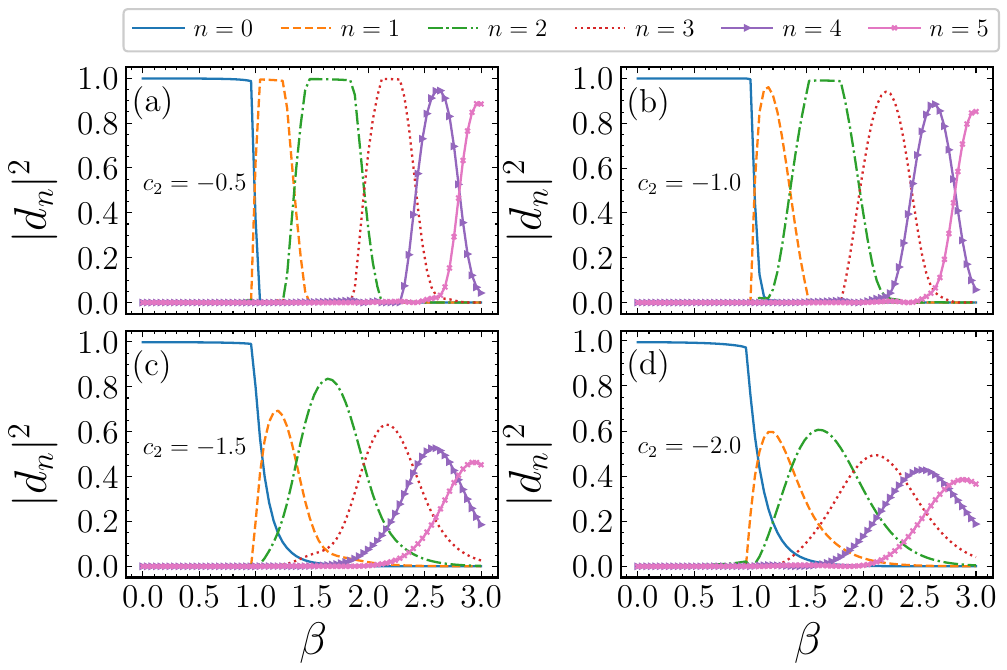}
\caption{The dependence of weights $|d_{n}|^{2}$ of linear eigenstates in
the GS expansion~\eqref{expand} on $\protect\beta $ for (a) $c_{2}=-0.5$,
(b) $c_{2}=-1.0$, (c) $c_{2}=-1.5$, (d) $c_{2}=-2.0$. The density-density
interaction remains repulsive, with $c_{0}=1$. As $c_2$ decreases, the
ground states of the system transitions from (a,b) mixed states formed by
the superposition of two adjacent energy levels to (c,d) edge states formed
by the superposition of a larger number of energy levels.}
\label{figure6}
\end{figure}

\begin{figure}[tbp]
\centering
\includegraphics[width=3.4in]{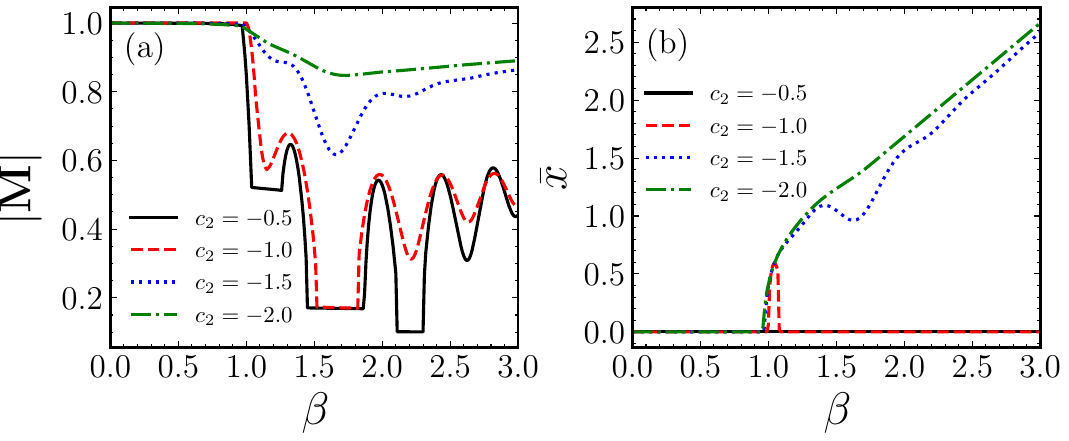}
\caption{(a) The absolute value of the magnetization vector and (b) the
average displacement $\bar{x}$ of the GS of the system with the spin-spin
attraction ($c_{2}<0$) vs. $\protect\beta $ for $c_{2}=-0.5,-1.0,-1.5$ and $%
-2.0$. The density-density interaction remains repulsive, with $c_{0}=1$. As
$c_2$ decreases, the absolute value of the magnetization gradually
increases. }
\label{figure7}
\end{figure}

Figure~\ref{figure6} shows the dependence of weights $|d_{n}|^{2}$ on $\beta
$ for $c_{2}=-0.5,-1,-1.5$ and $-2$. First, it is clearly seen that, for $%
\beta <1$, under the action of the spin-spin attraction ($c_{2}<0$), the GS
is virtually tantamount to the linear eigenstate with $n=0$, the entire
structure of the decomposition being nearly identical for $c_{2}=-0.5$ and $%
-1$. Near the critical point $\beta _{n}$ of the linear system, mixed states
emerge, which are chiefly formed as a superposion of eigenstates with
quantum numbers $n$ and $n+1$, with the weights satisfying $%
|d_{n}|^{2}+|d_{n+1}|^{2}\approx 1$. Far from the critical point, the GS
reverts to a shape which is close to the respective linear eigenstate, as
seen in Figs.~\ref{figure6}(a) and (b). Thus, we conclude that, in the case
of the weak spin-spin attraction, \textit{viz}., for $0\leq -c_{2}\leq 1$,
the GS phase transition remains close to the one in the linear system, but
between the mixed states, which are superpositions of states $\psi ^{(n)}$
and $\psi ^{(n+1)}$. As the attractive interaction strengthens, \textit{viz}%
., in the cases of $c_{2}=-1.5$ and $-2$, the system's GS becomes narrow
edge (spatially shifted) states, which are, essentially, superpositions of
several linear eigenstates, as seen in Figs.~\ref{figure6}(c) and (d).

The mixed states, which are formed, approximately, by the superposition of
two eigenstates, feature phase difference $\pi /2$ between the respective
weights $d_{n}$ and $d_{n+1}$ (see Eq. (\ref{expand})). Accordingly, one may
consider $d_{n}$ as a real number, while $d_{n+1}\approx i\sqrt{1-d_{n}^{2}}$%
. Then, the wave function of the mixed state is written as
\begin{equation}
\psi (x)=d_{n}\psi ^{(n)}(x)+i\sqrt{1-d_{n}^{2}}\psi ^{(n+1)}(x),
\label{psi_mix}
\end{equation}%
where the eigenfunctions $\psi ^{(n)}$ are defined in Eq.~\eqref{eigenstate}
and Eq.~\eqref{eigenstate2}. The magnetization (\ref{M}) corresponding to
this wave function is
\begin{equation}
M_{x}=0,\quad M_{y}=\frac{d_{n}\sqrt{1-d_{n}^{2}}}{\sqrt{2\left(
2-1/n\right) }}+\frac{d_{n}\sqrt{1-d_{n}^{2}}}{\sqrt{2\left( 2+1/n\right) }},
\label{My}
\end{equation}
\begin{equation}
M_{z}=\frac{d_{n}^{2}}{4n-2}+\frac{1-d_{n}^{2}}{4n+2}.  \label{Mzs}
\end{equation}%
Note that, unlike the linear eigenstates, the $y$-component of the
magnetization of the mixed state does not vanish. It attains a maximum in
the case when the two weights are equal, $d_{n}^{2}=d_{n+1}^{2}=1/2$.

Figure~\ref{figure7}(a) shows the dependence of the absolute value of the
magnetization, $|\mathbf{M}|$, on $\beta $ for different values of the
spin-spin-attraction coefficient $c_{2}$. First, we find that, for the same $%
\beta $, larger $|c_{2}|$ pushes $|\mathbf{M}|$ closer to the limit value $1$%
. For the mixed states corresponding to $c_{2}=-0.5$ and $c_{2}=-1.0$, the $|%
\mathbf{M}|\left( \beta \right) $ curves include fragments downward-opening
parabolas, which can be explained by the consideration of Eq. \eqref{My}.

The formation of the edge (spatially-shifted) GS can be analyzed in terms of
energy. To minimize the energy induced by spin-spin interactions, as given
by Eq.~\eqref{Es} with $c_{2}<0$, the local spin vector tends to have $|%
\mathbf{S}|=1$. Therefore, the respective GS tends to become a ferromagnetic
state. Typical spinors corresponding to such states are
\begin{equation}
\xi _{3}=\left[
\begin{array}{c}
1 \\
0 \\
0%
\end{array}%
\right] ,\xi _{4}=\frac{1}{2}\left[
\begin{array}{c}
1 \\
\sqrt{2} \\
1%
\end{array}%
\right] ,  \label{xi2}
\end{equation}%
cf. expressions (\ref{xi}) for the polar states. The state corresponding to $%
\xi _{3}$ implies that all atoms are collected in the $\psi _{1}$ component,
the same as in the linear solution~\eqref{eigenstate2} with $n=0$. This is
the reason why the spin-spin interaction does not affect states with quantum
number $n=0$ at $\beta \leq 1$, as shown in Figs.~\ref{figure5} and \ref%
{figure6}.

The GS corresponding to spinor $\xi _{4}$ in Eq. (\ref{xi2}), indicates that
the wave functions of the three components are related \ by
\begin{equation}
\psi _{0}(x)=\sqrt{2}\psi _{1}(x)=\sqrt{2}\psi _{-1}(x).  \label{proportion}
\end{equation}%
In this case, the spin-orbit coupling and Zeeman splitting can be neglected
because their energies,
\begin{equation}
\begin{aligned} E_{SO}&=i\beta\int_{-\infty}^{\infty}\psi^\dagger
F_y\partial_x \psi dx\\ &=\frac{\beta}{\sqrt{2}}\int_{-\infty}^{\infty}
\left[(\psi^*_1-\psi^*_{-1})
\partial_x\psi_0+\psi^*_0\partial_x(\psi_{-1}-\psi_1)\right]dx\\
\end{aligned}
\end{equation}
and
\begin{equation}
E_{Z}=\Omega\int_{-\infty}^{\infty}|\psi_1|^2-|\psi_{-1}|^2 dx
\end{equation}
are vanishingly small. By substituting these approximations in Eq. (\ref%
{submain}), we reduce it to a single-component equation,
\begin{equation}
\begin{aligned}
\mu\psi_0=-\frac{1}{2}\partial_x^2\psi_0+\left(\frac{x^2}{2}-\beta
x\right)\psi_0+ 2c_0|\psi_0|^2\psi_0, \end{aligned}
\end{equation}%
which features an effective potential $\overline{V}(x)=x^{2}/2-\beta x$.
Thus the atoms tend to accommodate around the obvious minimum of the
potential, $x=\beta $, which is shifted off the central position, as seen in
Figs.~\ref{figure5}($c_{1-3}$) and ($d_{1-3}$).

\section{The ground state of the system in the presence of the quadratic
Zeeman shift}

The above discussions focused only on the linear Zeeman effect; however, in
actual experiments, the quadratic Zeeman shift can have a significant impact
on the results. Incorporating the quadratic Zeeman effect, with strength $q$%
, into Eq.~\eqref{main} by introducing the respective Hamiltonian term $V_{%
\text{QZ}}=q(\mathbf{B}\cdot \mathbf{F})^{2}$, the modified Gross-Pitaevskii
equation is written as ~\cite{wenlin,QZ}
\begin{equation}
\begin{split}
i\partial _{t}\Psi =& \left[ -\frac{\partial _{x}^{2}}{2}+\frac{x^{2}}{2}%
+i\beta F_{y}\partial _{x}+\mathbf{B}\cdot \mathbf{F}+q(\mathbf{B}\cdot
\mathbf{F})^{2}\right.  \\
& +c_{0}\rho +c_{2}\rho \mathbf{S}\mathbf{\cdot }\mathbf{F}\bigg]\Psi ,
\end{split}
\label{main2}
\end{equation}%
where the magnetic field is the same as above, \textit{viz}., $B=-(\beta
x,0,1)$. To eliminate the influence of spin-spin interactions on the
system's GS, we set $c_{2}=0$ in this section, while the density-density
interaction is kept fixed at $c_{0}=1$.

\begin{figure}[bp]
\centering
\includegraphics[width=3.4in]{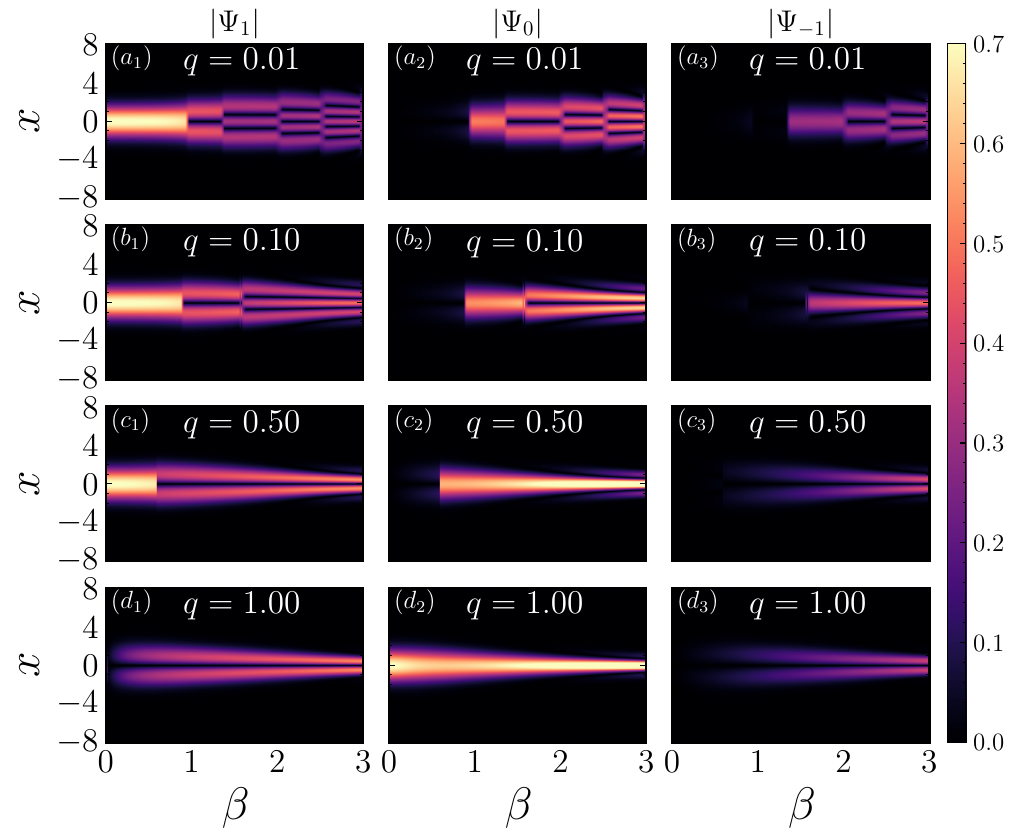}
\caption{Distributions of the absolute values of the wave functions in the
three GS components, in the presence of the quadratic Zeeman effect with
strength $q$ [see Eq. (\protect\ref{main2})], for ($a_{1-3}$) $q=0.01$, ($%
b_{1-3}$) $q=0.10$, ($c_{1-3}$) $q=0.50$, and ($d_{1-3}$) $q=1.00$. The
density-density and spin-spin interactions are set as $c_{0}=1$ and $c_{2}=0$%
, respectively.}
\label{figure8}
\end{figure}

We employed the imaginary-time evolution method to produce the system's GS
for different values of $q$. The numerical results are presented in Fig.~\ref%
{figure8}. For small values of $q$, such as $q=0.01$, the GS phase
transitions still occurs, being very similar to those reported above.
However, as $q$ increases, specifically at $q=0.1$, only a few lowest GS
phase transitions take place. We also observe that, as $\beta $ increases,
the range of the wave function distribution in the $x$ direction narrows.

For $q\geq 0.5$, the quadratic Zeeman effect is the dominant factor in the
system. Thus, the distribution of the GS wave function can be determined by
analyzing the energy induced by the quadratic Zeeman effect, with the energy
expressed as
\begin{equation}
E_{\text{QZ}}=q\int_{-\infty }^{+\infty }\psi ^{\dagger }\left( \mathbf{B}%
\cdot \mathbf{F}\right) ^{2}\psi dx.  \label{EQZ}
\end{equation}%
We further simplify Eq.~\eqref{EQZ} by introducing the unitary operator
\begin{equation}
U=\exp (i\theta F_{y})=I+F_{y}^{2}(\cos \theta -1)+iF_{y}\sin \theta ,
\label{U}
\end{equation}%
where $I$  is the identity matrix of size $3\times3$. The physical meaning of this operator is the
clockwise rotation by an angle $\theta $ around the $y$-axis, which makes
the magnetic field pointing in the $-z$ direction. The rotation angle $%
\theta $ is determined by expressions
\begin{equation}
\cos \theta =\frac{1}{\sqrt{\beta ^{2}x^{2}+1}},\quad \sin \theta =\frac{%
\beta x}{\sqrt{\beta ^{2}x^{2}+1}}.  \label{sintheta}
\end{equation}%
Till now, it is easy to check $U\mathbf{B}\cdot \mathbf{F}U^{\dagger }=-|%
\mathbf{B}|F_{z}=-\sqrt{\beta ^{2}x^{2}+1}F_{z}$. With the aid of the
unitary operator~\eqref{U}, the energy~\eqref{EQZ} can be simplified to:
\begin{equation}
E_{QZ}=q\int_{-\infty }^{\infty }(\beta ^{2}x^{2}+1)\rho ^{2}\xi ^{\dagger
}U^{\dagger }\cdot F_{z}^{2}U\xi dx.  \label{Eqz}
\end{equation}%
It is evident that an effective potential $\widetilde{V}=q\beta ^{2}x^{2}$
emerges, which corresponds to the harmonic-oscillator one. As $\beta $
increases, the strength of the potential grows, leading to a narrowing of
the wavefunction distribution along the $x$-direction.

Being concerned with the GS of the system, we minimize energy $E_{\text{QZ}}$%
. It is important to note that, when $U\xi =(0,1,0)^{T}$, the energy reaches
its minimum, which is $E_{\text{QZ}}=0$. Thus, we obtain the corresponding
spinor
\begin{equation}
\xi _{\text{GS}}=U^{\dagger }\left[
\begin{array}{c}
0 \\
1 \\
0%
\end{array}%
\right] .  \label{xi3}
\end{equation}%
Next, we consider two essential spatial regions. The first one corresponds
to $x=0$, where we have $\cos \theta =1$ and $\sin \theta =0$, obtaining $U=I
$ and $\xi _{\text{GS}}=(0,1,0)^{T}$. The second region corresponds to $%
\beta x\gg 1$, where $\cos \theta =0$ and $\sin \theta =1$, resulting in $%
U=I-F_{y}^{2}+iF_{y}$ and $\xi _{\text{GS}}=\left(1/\sqrt{2}\right)(1,0,1)^{T}$. These
conclusions precisely explain the wavefunction distribution patterns
observed in Figs.~\ref{figure8}($c_{1-3}$) and ($d_{1-3}$)). In the former
figure, the GS transition occurs at $\beta =0.6$, and the pattern beyond
this point is similar to the one in Fig.~\ref{figure8}($d_{1-3}$). We
conclude that the particles predominantly occupy component $\psi _{0}$
around $x=0$, corresponding to $\xi _{\text{GS}}=(0,1,0)$. As $x^{2}$
gradually increases, the particles are transferred to components $\psi _{\pm
1}$, until full occupancy is achieved, corresponding to the spinor 
$\xi _{\text{GS}}=\left(1/\sqrt{2}\right)(1,0,1)^{T}$.

\section{Conclusion}

This work is focused on the phase transition in the GS (ground state) in
spin-1 BECs, considering the combined effects of SOC (spin-orbit coupling)
and gradient magnetic fields. The linear system is solved exactly by means
of the formulation using raising and lowering operators. Through the
analysis of the energy spectrum, we have demonstrated that, following the
variation of the magnetic-field gradient and SOC strength, the system
undergoes a series \ of GS phase transitions, allowing all excited states to
transition into the GS.

The nonlinear system, including the density-density and spin-spin
interactions, has been solved numerically. In the case of the spin-spin
repulsion, the results obtained for the nonlinear system are similar to
their counterpart for the linear one. The results are drastically different
in the case of the spin-spin attraction: the relatively weak interaction
gives rise to bimodal mixed states near the GS phase-transition points of
the linear system, while stronger interaction creates the narrow GS as the
edge state shifted off the central position. The latter finding is explained
by means of the analytical consideration. A natural continuation of the
current work is to consider those problems in 2D, where we can explore the
GS phase transitions of vortex states.

\section*{Acknowledgments}

This work was supported by the GuangDong Basic and Applied Basic Research
Foundation through grant No. 2023A1515110198, Natural Science Foundation of
Guangdong Province through grants Nos. 2024A1515030131 and 2021A1515010214,
Foundation for Distinguished Young Talents in Higher Education of Guangdong No. 2024KQNCX150,
National Natural Science Foundation of China through grants Nos. 12274077, 62405054,
12475014 and 11905032, the Research Found of the Guangdong-Hong Kong-Macao Joint
Laboratory for Intelligent Micro-Nano Optoelectronic Technology through
grant No. 2020B1212030010, and Israel Science Foundation through Grant No.
1695/22.

\end{document}